\documentclass{iau}
\usepackage{graphicx}
\title[JD 11.~~Supernovae and the Galactic Ecosystem] 
{Supernovae and the Galactic Ecosystem}

\author[]{Q. Daniel Wang$^1$}

\affiliation{$^1$Astronomy Department, \\ 
619-E, LGRT,  University of Massachusetts,   710 N. Pleasant St.,
Amherst, MA 01003, USA \\
 email: {\tt wqd@astro.umass.edu} }

\pubyear{2013}
\volume{296}  
\pagerange{119--126}
\jname{IAU 296 Supernova environmental impacts}
\editors{Richard McCray, Alak Ray}
\begin{document}

\maketitle

\begin{abstract}
Supernovae are the dominant source of stellar feedback, which plays an important role in regulating galaxy formation and evolution. While this feedback process is still quite uncertain, it is probably not due to individual supernova remnants as commonly observed. Most
supernovae likely take place in low-density, hot gaseous environments,
such as superbubbles and galactic bulges, and typically produce no
long-lasting bright remnants. I review recent observational and
theoretical work on the impact of such supernovae on galaxy ecosystems, particularly on hot gas in superbubbles and galactic spheroids. 
\end{abstract}

\firstsection 
\section{Introduction}

Supernovae (SNe) are a major source of the mechanical energy input in
the  interstellar medium (ISM). On average, an SN releases about
$10^{51}{\rm~ergs}$ energy, driving a blastwave
into the ambient medium. How far  this blastwave goes and how fast
the energy is dissipated depend sensitively on the density and temperature of
the medium. Ironically, commonly-known and well-studied supernova remnants
(SNRs), though looking spectacular, are atypical products of SNe,  and
typically occur in relatively dense media. Most SNe are expected to explode in
low-density hot environments, e.g., inside superbubbles (SBs) and/or simply in the
inter-cloud hot ISM (for core-collaped SNe) and in the
Galactic halo and bulge (for Type Ia SNe). The resultant remnants are
typically too faint to be well observed individually. But collectively such
``missing'' SNRs are probably more important than those in dense environments,
in terms of both heating and shaping the global ISM. In the following,
I will first briefly review the behavior of SN blastwaves in hot gas and
will then focus on discussing how discrete SN events affect hot gas properties in the two types
of environments, SBs and galactic spheroids.
 \firstsection 

\section{Supernova blastwaves in hot gas}

As shown by \cite{Tang05}, the evolution of an SNR in a low-density hot medium
has several distinct properties: 1) The blastwave always moves 
at a speed greater than, or comparable to, the sound wave and can thus reach 
a much larger radius than that predicted
by the Sedov solution; 
2) The swept-up thermal energy is important, affecting the evolution of both the blastwave and
the interior structure; 3) Because the Mach number of the blastwave is typically small, its heating
is subtle and over a large volume; 4) The blastwave can hardly
dissipate until it meets cool gas.  These properties of SNRs in a
low-density hot medium make them an important ingredient in regulating
large-scale environments in galaxies. 

To study the collective effect of SNe on the ISM, we need to conduct
simulations, which needs to cover a large dynamic range from the evolution
of individual SNRs to a possible galaxy-wide outflow, for example. It
would be computationally very expensive, if even possible, to simulate
the evolution of each SNR on sub-parsec scales in such a simulation. The SNR evolution in general cannot
be described by the self-similar Sedov–Taylor solution, which
neglects the SN ejecta and assumes a cool ambient medium (hence
with no energy content). In fact, the evolution depends on both the
density and temperature of the ambient medium (Tang \& Wang
2005). Tang \& Wang (2009) have further shown that one can adaptively embed
individual structured SNR seeds into the 3D simulation grid on
such scales that their thermal and chemical evolution are adequately
represented by detailed 1D simulations. The SNR seed embedding,
worked with an adaptive mesh refinement scheme, can effectively
extend such a 3D simulation to include the subgrid evolution of
SNRs. In fact, using a scaling law, which is applicable to SNRs
evolving in hot gas, individual seeds can be adaptively generated from a library of templates. Each consists
of the radial profiles of density, temperature and velocity when
the SNR has a certain shock front radius or age. These templates
can be obtained from a single 1-D simulation of an SNR evolving in a
uniform ambient medium of certain density and temperature.
With this scheme, 3D hydrodynamic simulations of the
supernova-dominated ISM can be readily conducted.

\firstsection 

\section{Superbubbles}

Core-collapsed SNe represent the end of massive stars, the bulk (if not all) of which form in
OB associations. The energy release from such an association,
highly correlated in space and time, has a great impact on the
surrounding medium. Initially, the energy release from the 
association is primarily in the form of intense ionizing radiation
from very massive stars, which tends to homogenize the surrounding
medium and to make it puff up. After several million years,
fast stellar winds start to play a major role in heating and shaping
the medium, creating a low-density hot bubble (e.g., \cite{Weaver77}), before the explosion of the first SN
in the association. Later, after about $5 \times 10^6$~yr since the star formation
(if more or less coeval), core-collapsed SNe become the dominant
source of the mechanical energy input into the already
hot surroundings (e.g., \cite{Monaco04}). This combination of the concerted
feedbacks, lasting for $\sim 5 \times 10^7$~yr --- the lifetime of an
8 $M_\odot$ star --- leads to the formation of a so-called superbubble of
low-density hot gas enclosed by a supershell of swept-up cool
gas (e.g., Mac Low \& McCray 1988). The expansion of such
a superbubble is expected to be substantially faster than typical
OB association internal velocities of a few kilometers per
second. Therefore, a majority of core-collapsed SNe ($\sim 90\%$)
should occur inside their parent superbubbles (e.g., Higdon
\etal\ 1998; Parizot \etal\ 2004).

\begin{figure*}[b]
\vspace*{-1.0 cm}
\begin{center}
 \includegraphics[width=1.0\textwidth]{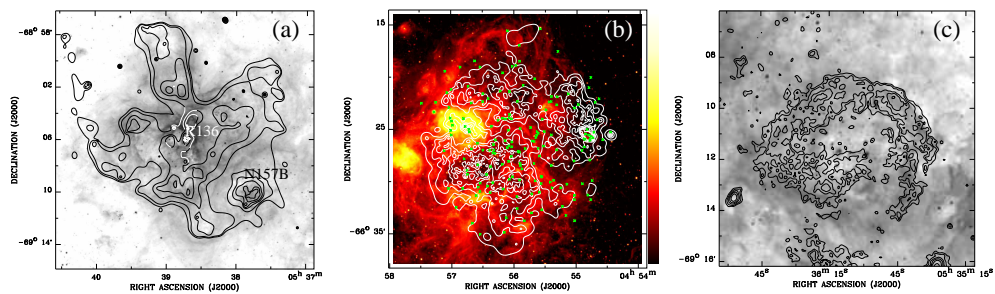} 
 \caption{Examples of superbubbles in the LMC: (a) 30 Dor, (b) N11,
   and (c)  30 Dor C. Overlaid  on the H$\alpha$ images are X-ray
   intensity contours constructed in the {\sl Chandra} ACIS-I 0.5-2
   keV band for (a) and (b) and in the {\it XMM-Newton} MOS 
0.4-1 keV band for (c). Discrete sources, as marked by{\sl crosses},
have been excised  in (b). }
   \label{fig:SB}
\end{center}
\end{figure*}

Fig.\,\ref{fig:SB} shows examples of SBs in various
evolutionary stages.
The 30 Dor nebula is probably the youngest and most energetic one,
while 30 Dor C is the oldest and least powerful. The diffuse X-ray-emitting plasma,
heated by fast stellar winds and possible SNe of massive stars, 
fills various cavities traced by H$\alpha$-emitting shells. Such
SBs will eventually blow out from galactic disks and vent
the hot plasma into galactic halos, even into the intergalactic space.

While the evolution model for SBs, analogous to that for stellar wind
bubbles, has been developed for many years (e.g., Weaver \etal\ 1977;  Mac Low \&
McCray 1988). Comparisons of this model with observations have revealed two major
discrepancies. The observed X-ray luminosities of some SBs seem to be substantially
brighter (by a factor of up to $\sim 10$) than predicted by the model
(Chu \& Mac Low 1990; Wang \& Helfand 1991; Jaskot \etal\ 2011 and
references therein). Various processes have been proposed,
including SN metal enrichment, additional  mass loading due to the destruction of embedded,
dense cloudlets of the ISM that have been overrun by expanding
supershells. But it appears that the most important process is the
blastwave heating of supershells by sporadic SNe,
especially off-center ones (\cite{Jaskot11} and references therein). 

The blastwave heating of supershells may also be responsible for 
another apparent discrepancy between the model prediction and the
observations: The energetics of SBs, observationally accounting for
both thermal and kinetic energies, is generally much less than expected from the injection from
massive stars enclosed (e.g., \cite{Yamaguchi10}). Or the growth rate
of an SB is typically much less than what would be expected from the model
(e.g., \cite{Oey09}). The energy injection from stars is assumed as
being continuous, which may be a reasonable approximation for an
evolved SB of a large dimension,  which the model was intended to
applyd to. But for a relatively young and small SB, as typically
observed, an SN-induced shockwave or sound wave can propagate through the hot
interior, eventually reach the surrounding cool shells, and
steepen again into a blastwave, somewhat like an earth quake-induced Tsunami
(\cite{Tang05}). This blastwave, or its corresponding reverse-shock,
can sometimes be strong enough to produce enhanced X-ray emission, as
observed. Furthermore, much of the SN energy may be lost radiatively,
probably  outside the X-ray band. 
Indeed, as shown by \cite{Cho08}, based on 1-D simulations of such
young SBs, discrete SNe can lead to a loss of 80-90\% of the SN mechanical energy via radiative cooling in
supershells, and only about 10\% remains as the thermal energy in the enclosed
hot gas, while the kinetic energy is $\lesssim 10\%$. Of course, such
a substantial energy loss can significantly change the growth of SBs, and
potentially their chance to blow out from galactic disks. Therefore,
it should be interesting to extend such simulations to higher dimensions 
and to later evolutionary stages of SBs. Interfaces between the hot
and cold gases can be enlarged, for example, due to various
3-D instabilities, which can potentially strongly affect the energetics of SBs (e.g., 
\cite{Krause13}).

In addition to X-ray emission, SBs may loss energy via other channels.
Substantial amounts of dust have been observed in various
Galactic wind-blown bubbles  (e.g., \cite{Everett10}). Dust can be a significant sink of
heat in hot gas. But the survival time of
dust in hot gas is short, due to sputtering. So dust needs to be
replenished, possibly by clouds overrun by
expanding shells and by dense stellar winds from massive stars such as
Carbon-rich Wolf-Rayets and Luminous Blue variables (e.g.,
{\cite{Rajagopal07, Dong12}). 

In addition to the radiation loss, SN energy may also
escape from SBs  via cosmic rays (CRs).
SBs  have been proposed to be the acceleration site of Galactic CRs
(e.g., \cite{Higdon05, Fisk12}; Parizot et al. 2004; Bykov 2001; Bykov
\& Fleishman 1992).  
The energy efficiency of the CR-acceleration in SBs could reach up to about 30\%
at their early evolutionary stages (ages of a few $10^6$ years; e.g., \cite{Butt08})
Nonthermal X-ray emission has indeed been observed in a number of
SBs, including 30 Dor C (\cite{Smith04, Yamaguchi10} 
 in the LMC, IC131 in M33 (\cite{Tullmann09}). Fig.~\ref{fig:SB_spec}
compares X-ray spectra of 30 Dor and 30 Dor C. While the former is dominated by
   thermal emission, the latter exhibits a substantially hard
   (apparently nonthermal) component (\cite{Smith04}).

\begin{figure*}[t]
\begin{center}
 \includegraphics[width=0.95\textwidth]{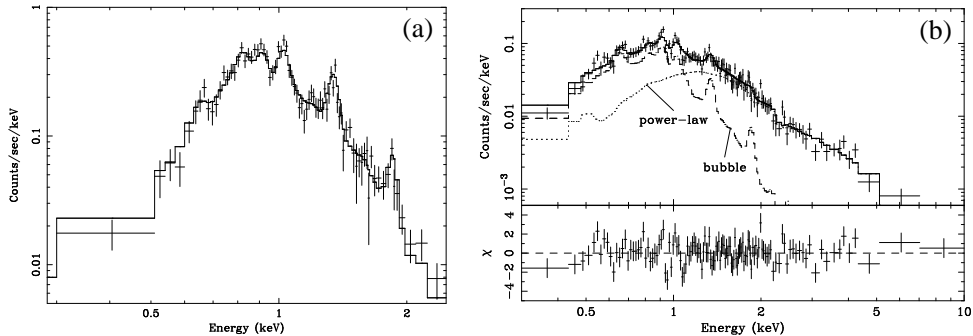} 
 \caption{(a) Comparison of XMM-Newton MOS spectra of (a) 30 Dor,
   excluding R136 and SNR N157B (Fig.~\ref{fig:SB}a) and  (b)
   30 Dor C. }
   \label{fig:SB_spec}
\end{center}
\end{figure*}

All these processes still need to be carefully studied to determine
their actual roles in regulating the energetics of SBs, which could
have strong implications for our understanding the impact of massive
stars on galaxy ecosystems.

\firstsection 

\section{Hot gas in galactic spheroids}

Spheroids, including bulges of spiral galaxies and elliptical galaxies, are primarily comprised of 
old stars, which account for more than half of the stellar mass in the local Universe 
(Fukugita, Hogan, \& Peebles 1998). These stars collectively generate a substantial feedback 
in form of gradual stellar mass-loss and energetic Type Ia SNe. Understanding how this relatively 
gentle but long-lasting feedback affects galaxy evolution is one of the fundamental questions in
astrophysics (e.g., Wang 2010). In order to address this issue, one
needs an affective tracer of the 
ejected mass and energy in spheroids. While they typically
contain little cool gas, one would expect that the mass should be
mostly in hot gas, heated by the SNe. Indeed, diffuse optically-thin
thermal X-ray emission has been observed in elliptical galaxies and in galactic bulges. However it has been shown repeatedly that the X-ray-inferred gas mass and 
energy are far less than those empirical predictions (e.g. David et
al. 2006), which becomes particularly acute in X-ray-faint spheroids
(bulges of spirals and low/intermediate-mass ellipticals). In such a
spheroid, the luminosity of the X-ray emission accounts for no more
than a few percent of the expected SNe energy input, and the deduced
iron metallicity is typically sub-solar, inconsistent with the expected Ia SN enrichment. The inferred total 
mass of the X-ray-emitting hot gas represents only a small fraction of
what is expected from the stellar mass loss over the galaxy's
lifetime. These discrepancies clearly indicate that the stellar feedback in such a spheroid
has gone with a wind or outflow (e.g., Bregman 1980; Ciotti et al. 1991; David et al. 2006; Tang et al. 2009a, 2009b; Tang \& Wang 2010). 

The study of hot gas in galactic spheroids has quite a long
history. Early works were based primarily on 1-D galactic wind
modeling. Except for massive ellipticals,  very hot ($>\!\!1\,\rm keV$) and fast ($v>800\,\rm km/sec$)
spheroidal wind was predicted (e.g., Mathews \& Baker 1971; Ciotti et
al. 1991). For such a hot wind, the inferred X-ray emission should
be much less than $10^{36}\,\rm ergs\,s^{-1}$ (e.g., Tang et
al. 2009b), which is two orders of magnitude lower than the recently
observed value of the galactic spheroid ($\sim 3\times 10^{38}\,\rm
ergs\,s^{-1}$ for the M31 bulge; Li \& Wang 2007; Bogdan \& Gilfanov 2008).

One important fact that the 1-D supersonic wind modeling does not
account for is the discreteness of Ia SN heating. X-ray emission,
proportional to the emission measure, is sensitive to the detailed
structure of the hot gas. We have therefore conducted 3-D simulations
of SN-driven galactic winds (Tang et al. 2009b), in which the energy
is assumed to come mainly from sporadic SNe and the mass mainly from
stellar winds. The results are illustrated in Fig.~\ref{F:bgwimage}. Our simulations confirm that Ia SNe produce a spheroidal wind,
and more importantly, reveal its substructures. The bulk of the X-ray emission originates from the relatively 
low-temperature and low-abundance gas shells associated with SN blastwaves; the SN ejecta are not well mixed with the ambient medium. These results are qualitatively consistent with the apparent lack of evidence for iron enrichment in X-ray-faint galaxies. Compared to the 1-D wind model, the non-uniformity of simulated gas density, temperature, and metallicity in the 3-D simulations 
increases the diffuse X-ray luminosity by a factor of a few, narrowing
the discrepancy between the theory-predicted and the observed X-ray emission. However, the resultant diffuse luminosities in the 3-D simulations is still more than one order of magnitude less than 
those observed in the Galactic and M31 bulges, indicating that gas in these spheroids is in a subsonic outflow state, probably due to additional mass loading to the hot gas and/or due to energy input rate that is substantially lower than the current estimate.

\begin{figure}[ht]
\vspace*{-0.5 cm}
\includegraphics[width=1\textwidth]{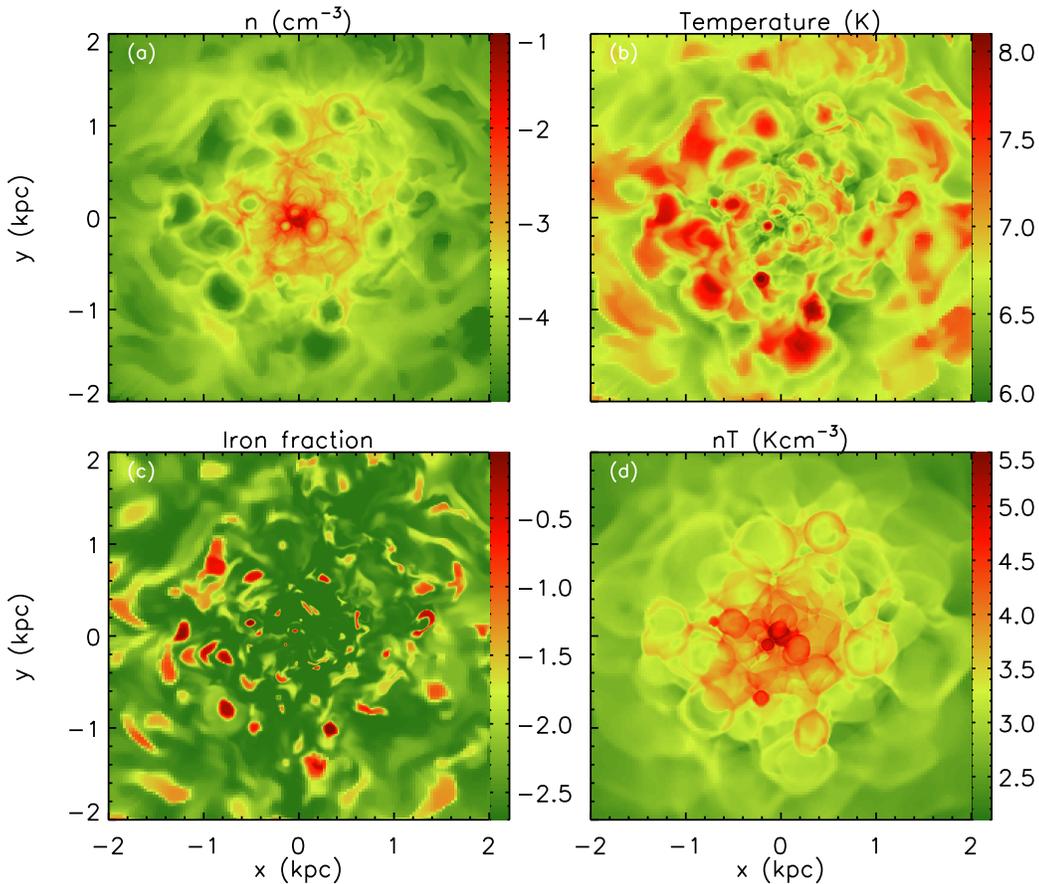}
\caption{\label{F:bgwimage}Density, temperature, iron mass fraction,
and pressure of gas in a cut through a simulated spheroidal wind. The
image values are scaled logarithmically.
The upper-right octant has a higher resolution in the
simulation, allowing for testing various resolution effects.}
\end{figure}

A subsonic outflow from a spheroid is intimately related to its
formation and evolution history (\cite{Ciotti91,Tang09a}). We have developed a 1-D model of the 
spheroid outflow and galaxy accretion interplay.
This model is based on 1-D hydrodynamic simulations and on an 
approximated dark matter accretion history of galaxies 
(Tang et al. 2009a).
In our model, the outer boundary condition is the (supersonic) Hubble flow,
which avoids the boundary problem.
The spheroid wind/outflow occurs well inside the spatial range of the simulations.
The feedback is assumed to consist of two primary phases:
1) an initial burst during the spheroid formation and 2) a subsequent
long-lasting mass and energy injection from stellar winds and Ia SNe
of evolved low-mass stars. These two phases of the feedback re-enforce 
each-other's impact
on the gas dynamics. An outward blastwave is initiated by the
burst and is maintained and enhanced by the long-lasting stellar feedback. 
For an M31-like bulge, for example, this blastwave 
can heat the surrounding medium not only in the galactic halo, 
but also in regions beyond the virial radius.  
The long-lasting feedback forms a galactic spheroid wind initially, which is 
reverse-shocked at a large radius, and may later evolve into a 
stable subsonic outflow as the energy injection gradually 
decreases with time. In a subsonic outflow, the properties of hot gas 
depend sensitively on the environment 
and formation history, which can explain the large dispersion of $L_X/L_K$ in
early-type galaxies with similar $L_K$ as well as
the missing stellar feedback problem (Wang 2010 and references therein). Therefore, the understanding of the interplay
between the hot gas outflows and the large-scale galaxy environment is essential
to the correct interpretation of observed diffuse X-ray emission in and 
around galactic spheroids.

\cite{Tang10} have further conducted 3-D simulations based on the 1-D
subsonic solution to explore combined effects. In addition to the
expected enhanced X-ray emission, they find that SN reverse shock-heated iron ejecta is typically found to have a very high temperature and low density, hence producing little X-ray emission. Such hot ejecta, driven by its large buoyancy, can quickly reach a substantially higher outward velocity than the ambient medium, which is dominated by mass-loss from evolved stars. The ejecta is gradually and dynamically mixed with the medium at large galactic radii. The ejecta is also slowly diluted and cooled by in situ mass injection from evolved stars. These processes together naturally result in the observed positive gradient in the average radial iron abundance distribution of the hot gas, even if mass weighted. This trend is in addition to the X-ray measurement bias that tends to underestimate the iron abundance for the hot gas with a temperature distribution. 

One limitation of our previous simulations is that the mass loss from evolved stars takes only the form of continuous input, following the average stellar distribution of a galactic spheroid.
In reality, however, the continuous wind, mainly from constant mass loss of numerous giant stars, contributes only about half of the 
stellar mass loss in the spheroid (e.g. Buckley \& Schneider 1995). 
The other half of mass loss occurs in the planetary nebula (PN) stage
of intermediate and low mass stars. This mass loss  is an approximately
impulsive mass-loss event at the end of the life of such a star and lasts only
$\sim 10^3$ years. The ejecta forms an expanding ($\sim 10$ km/s)
and cold ($\lesssim 10^5$ K) gas shell. Naturally, one wants to know
whether or not the PN ejecta can be heated by shocks and/or by
thermal mixing with the hot ambient gas, or how the discreteness of
the mass input can affect the diffuse X-ray emission of a spheroid.

Bregman \& Parroit (2009) have conducted 2-D hydrodynamic simulations of an
individual PN undergoing ram-pressure stripping of ambient hot gas in a
typical spheroid environment. They show that fluid instabilities result
in mixing and heating of about half of the PN ejecta to $10^5-10^6$ K,
a temperature range that is still at least a factor of several lower
than that of the hot gas, while the other half remains cold ($<10^5$
K) and flows out of the simulation box (25 pc long in the opposite
direction of the motion). These results provide useful insights into the
PNe feedback processes. But clearly the simulations have several major
limitations, in terms of the explored volume, strength/mode of fluid
instabilities,  and interplay among various objects/processes in a
spheroid.

In a real galactic spheroid, the environment can be very violent because of the 
presence of SN shockwaves and the collisions among randomly moving
PNe, as well as the global spheroid-wide outflow, and hence the large
pressure and density gradients. We have recently performed a pilot study, which
for the first time simulates a galactic spheroid with discrete PN
feedback. While the basic setting is the same as that described in
\cite{Tang10}, we make the PN mass contribution discrete, after the
SN-driven galactic outflow becomes stable (Tang \& Wang 2010). 
The positions and velocities of the imbedded PN seeds, with an occurrence rate of $4.5\times 10^4 \rm~Myr^{-1}$, are determined in a separate N-body simulation
using the software package Zeno (developed by Joshua Barnes). 
Each embedded PN seed has a mass of $0.3\rm~M_\odot$, an expansion
velocity of 20 $\rm~km~s^{-1}$,  and a uniform temperature of $10^5$ K
(artificially set to achieve a sufficiently high initial resolution). The size
of the seed is determined by a rough local pressure balance at the embedded position.
The refinement level reaches down to $\sim 1$ pc. The embedded PNe then continue to expand and interact with their environments. 

\begin{figure}[ht]
\vspace*{-0.5 cm}
\includegraphics[width=0.5\textwidth]{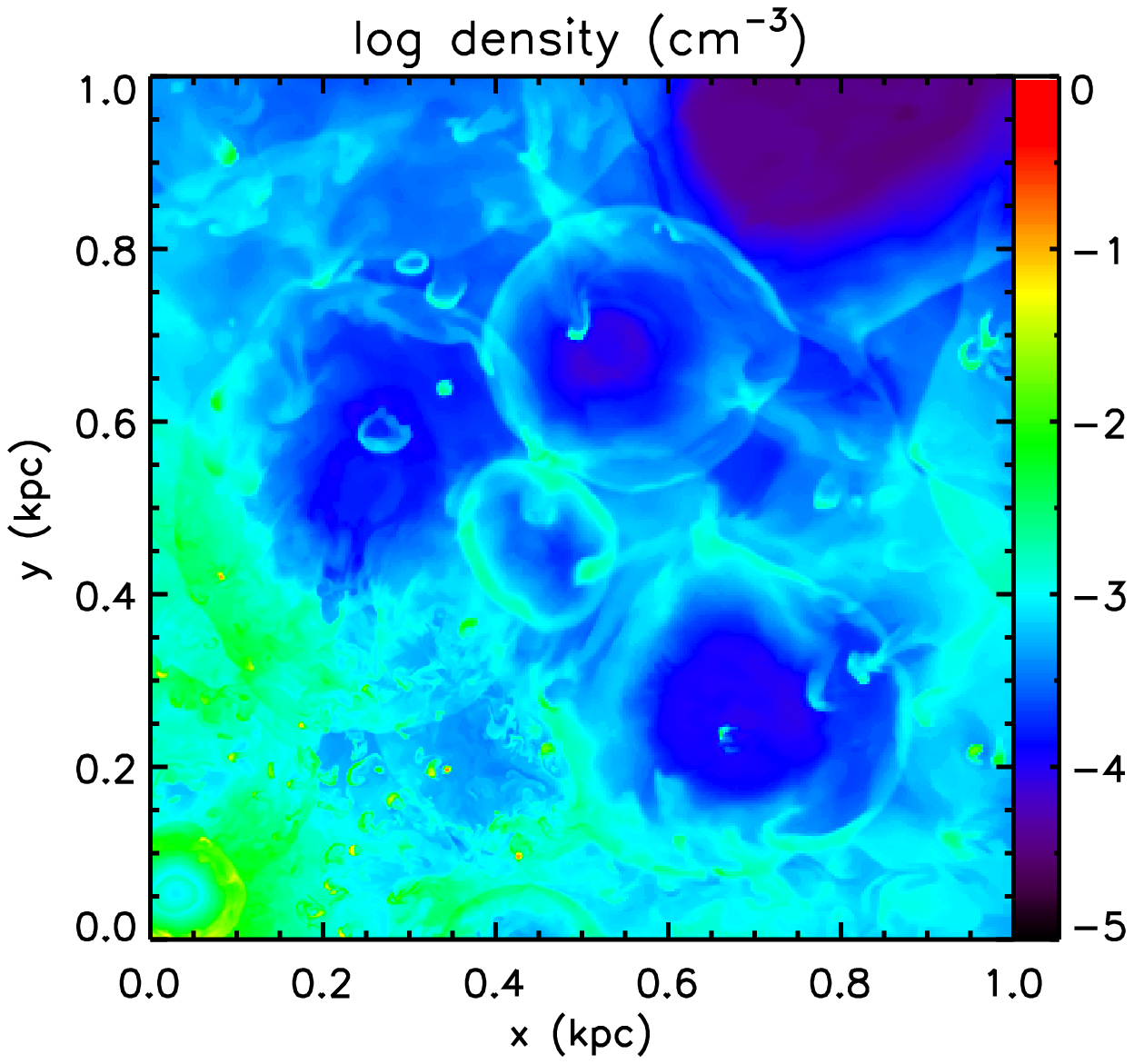}
\includegraphics[width=0.5\textwidth]{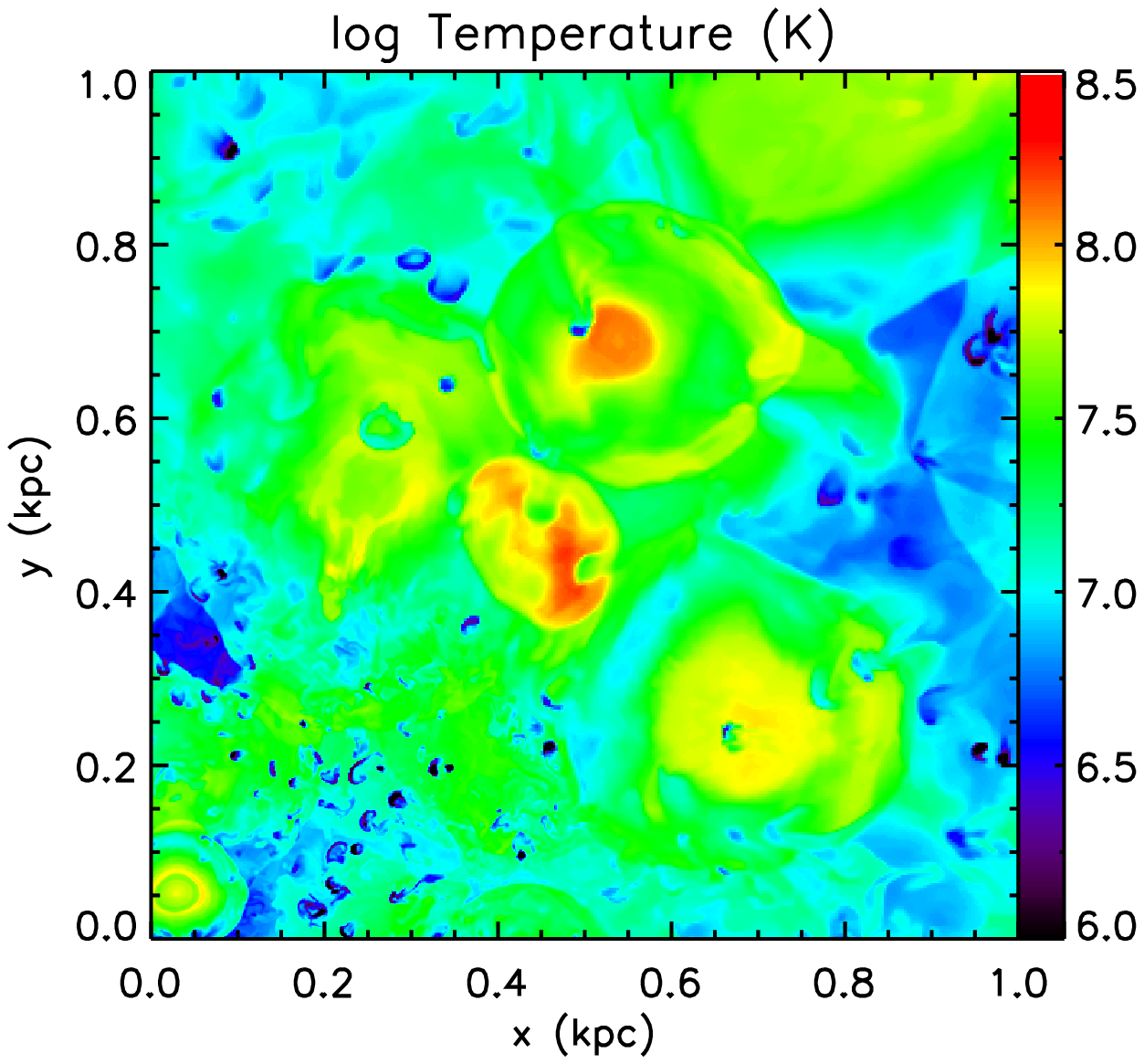}
\caption{A snapshot of our pilot simulation, including discrete PNe. Presented here are
  density (left) and temperature (right), viewed in a cut through  the 
central $1\rm kpc \times 1\rm kpc$ region of a spheroid. The simulation, which includes all 
three types of feedback (SNe + stellar winds + PNe), has evolved 1.3
Myr after discrete PN seeds start to be planted.}
\label{pne}
\end{figure}

Fig. \ref{pne} illustrates the result of the simulation, showing a diversity of PN morphologies on scales of a few pc to a few tens of pc, while SN shock waves are typically on larger scales. 
Close examination of individual PNe suggests that some of them are morphologically similar to that in 
the 2-D simulations (Bregman \& Parroit 2009), but others appear to be very different, 
demonstrating complicated interaction of PNe with the environments. 
In the inner most region where the gas density and stellar number
density are high, PNe only grow to several pc in size, while in outer
regions or inside low-density SNRs PNe can grow as big as 50 pc. Most interestingly, 
many of the PNe seem to be heated eventually to a couple of million
degrees, at which the bulk of the observed diffuse X-ray luminosity is
emitted. Therefore, the debris of PNe can naturally lead to enhanced
X-ray emission. Detailed simulation and analysis will enable us to examine the evolution of individual PNe to see how
instabilities grow in different local environments, to quantify what fraction of
PN mass is heated to a temperature high enough to emit X-ray, and to
check whether or not the enhanced X-ray emission could
make up the difference between previous simulation predicted value and the observed one.
 
In conclusion, most SNRs are not observed individually. These
``missing'' SNRs collectively represent a key component of the
stellar feedback in galaxies, in either active star-forming spirals or
cool gas-poor spheroids. The feedback regulates the ecosystems of the
galaxies, hence their evolution.

\firstsection 

\end{document}